\documentclass{aa}  
\usepackage{graphicx}
\usepackage{subfigure}
\usepackage{txfonts}
\usepackage{verbatim}
\usepackage{float}
\usepackage{pifont}
\usepackage{tabularx,adjustbox,booktabs}
\begin{document}

\title{Pulsation properties of ultra-massive DA white dwarf stars
  with ONe cores}

\author{Francisco C. De Ger\'onimo\inst{1,2}, Alejandro H. C\'orsico\inst{1,2},
  Leandro G. Althaus\inst{1,2}, Felipe C. Wachlin\inst{1,2}, Mar\'ia E. Camisassa\inst{1,2}}
\institute{$^1$Grupo de Evoluci\'on Estelar y Pulsaciones. Facultad de 
           Ciencias Astron\'omicas y Geof\'{\i}sicas, 
           Universidad Nacional de La Plata, 
           Paseo del Bosque s/n, 
           (1900) La Plata, 
           Argentina\\
       $^2$Instituto de Astrof\'{\i}sica La Plata, 
           IALP (CCT La Plata), 
           CONICET-UNLP\\
\email{fdegeronimo@fcaglp.unlp.edu.ar}}
\date{Received ; accepted }

\abstract{ Ultra-massive hydrogen-rich white dwarf stars are expected
  to harbor oxygen/neon cores resulting from the progenitor evolution
  through the super-asymptotic giant branch phase. As evolution
  proceeds during the white dwarf cooling phase, a crystallization
  process resulting from Coulomb interactions in very dense plasmas is
  expected to occur, leading to the formation of a highly crystallized
  core. In particular, pulsating ultra-massive white dwarfs offer a
  unique opportunity to infer and test the occurrence of
  crystallization in white dwarf interiors as well as physical
  processes related with dense plasmas.}{We aim to assess the
  adiabatic pulsation properties of ultra-massive hydrogen-rich white
  dwarfs with oxygen/neon cores.}{ We studied the pulsation properties
  of ultra-massive hydrogen-rich white dwarf stars with oxygen/neon
  cores. We employed a new set of ultra-massive white dwarf
  evolutionary sequences of models with stellar masses in the range
  $1.10 \leq M_{\star}/M_{\sun} \leq 1.29$ computed by taking into
  account the complete evolution of the progenitor stars and the white
  dwarf stage. During the white dwarf cooling phase, we considered
  element diffusion. When crystallization set on in our models, we
  took into account latent heat release and also the expected changes
  in the core chemical composition that are due to phase separation
  according to a phase diagram suitable for oxygen and neon
  plasmas. We computed nonradial pulsation $g$-modes of our sequences
  of models at the ZZ Ceti phase by taking into account a solid
  core. We explored the impact of crystallization on their pulsation
  properties, in particular, the structure of the period spectrum and
  the distribution of the period spacings.} {We find that it would
  be  possible, in principle, to discern whether a white dwarf has a
  nucleus made of carbon and oxygen or a nucleus of oxygen and neon by
  studying the spacing between periods.}{ The features found in the
  period-spacing diagrams could be used as a seismological tool to
  discern the core composition  of ultra-massive ZZ Ceti stars,
  something that should be complemented with detailed asteroseismic
  analysis using the individual observed periods.}

\authorrunning{De Gerónimo et al.}
\titlerunning{Pulsating massive white dwarfs}
\maketitle
%_____________________________________________________________________

\section{Introduction}

%%%%%%%%%%%%%%%% INTRO %%%%%%%%%%%%%%%%%%%%%%%%%%%%%%%%%%%%%%%%%%%%%%%%%%%%%%%%%%%%%%

White dwarf (WD) stars are the final stage of evolution for low- and
intermediate-mass stars. Most of these stars are post-asymptotic giant
branch (AGB) remnants with carbon-oxygen cores and stellar masses
$M_{\rm wd} \sim 0.6 M_{\sun}$. The mass distribution also exhibits a
tail of massive WDs peaking at $\sim 0.82 M_{\sun}$,
\citep[][]{2013ApJS..204....5K}. The existence of massive WDs with
$M_{\rm wd} \gtrsim 0.8 M_{\sun}$
\citep{2010MNRAS.405.2561C,2013MNRAS.430...50C} and ultra-massive WDs
with $M_{\rm wd} \gtrsim 1.15M_{\sun}$,
\citep{2013ApJ...771L...2H,2017MNRAS.468..239C} is also revealed.

Theoretical studies suggest that progenitor stars that are initially
more massive than $\sim 7 M_{\sun}$ are expected to evolve through the
super-AGB (S-AGB) phase, igniting carbon in their interiors and
eventually resulting in the formation of oxygen and neon (ONe)  WDs
with masses $M_{\rm wd}\gtrsim 1.05M_{\sun}$
\citep{1997ApJ...485..765G}. Specifically, carbon is ignited in
semi-degenerate conditions, leading to repeated carbon-burning shell
flashes that propagate inward, turning the CO core into an ONe mixture
\citep{1994ApJ...434..306G,2006A&A...448..717S}. After carbon is
exhausted in the core, and as a result of mass loss, the remnant
evolves toward the WD domain, and a WD is formed with a core made
mostly of $^{16}$O and $^{20}$Ne with traces of $^{23}$Na and
$^{24}$Mg \citep{2007A&A...476..893S}.

As is well known, many WDs exhibit multiperiodic luminosity variations
caused by pulsations. In particular, variable H-rich (DA) WDs, called
ZZ Ceti (or DAV), are the most numerous class of compact
pulsators. These stars show variations in their luminosity in a narrow
instability strip of $10,500 \lesssim T_{\rm eff} \lesssim 12,300$ K
\citep{2008PASP..120.1043F,2008ARA&A..46..157W,2010A&ARv..18..471A}
with photometric variations of up to 0.3 mag. These variations are the
result of spheroidal non-radial $g$-mode pulsations with low harmonic
degree ($\ell \leq 2$) and periods in the range 70-1500 s. In
particular, the first attempt of studying the adiabatic pulsational
properties of ultra-massive ONe-core WDs was made by
\citet{2004A&A...427..923C}, who demonstrated that pulsational
expectations such as the forward and mean period spacing of ONe WDs
are markedly different from those of CO WDs.

One interesting point in studying ultra-massive pulsating DA WDs lies
in the fact that these stars are expected to harbor a crystallized
core resulting from Coulomb interactions in very dense
plasmas. Although the occurrence of crystallization in WDs was
theoretically suggested several decades ago by \citet{Kirzhnits1960},
\citet{Abrikosov1961}, \citet{1961ApJ...134..669S} and
\citet{1968ApJ...151..227V} \citep[see also recent studies
  by][]{1999ApJ...526..976M,2004ApJ...605L.133M,2005A&A...429..277C,
  2005ApJ...622..572B}, it was not until quite recently that the
existence of crystallized WDs was inferred from the study of WD
luminosity function of stellar clusters
\citep{2009ApJ...693L...6W,2010Natur.465..194G}. The fact that
ultra-massive pulsating DA WDs are expected to be crystallized turns
these stars into unique objects from which the occurrence of
crystallization in WD interiors can also be inferred. The first star
studied in this sense was the ultra-massive ZZ Ceti star BPM 37093
\citep {1992ApJ...390L..89K,2005A&A...432..219K} , which is expected
to have a partially crystallized core
\citep{2004ApJ...605L.133M,2005ApJ...622..572B}.

We here aim to assess the adiabatic pulsation properties of
ultra-massive H-rich WDs with ONe cores on the  basis of full
evolutionary models that incorporate the most updated physical
ingredients governing the progenitor and WD evolution. This
investigation constitutes a substantial improvement over the study of
\citet{2004A&A...427..923C} in three major aspects: first, the
chemical profiles for all of our WD models,  which were taken from
\citet{2010A&A...512A..10S}, are consistent with  the predictions of
the progenitor evolution with stellar masses in the range $9.0 <
M_{\rm ZAMS}/M_{\sun} < 10.5$ from the zero-age main sequence (ZAMS)
to the end of the  the S-AGB phase. Thus, not only a realistic ONe
inner distribution expected for each WD mass is considered in the WD
modeling, but also realistic chemical profiles and inter-shell masses
built up during the S-AGB are taken into account. Second, we take into
account for the first time the changes in the core chemical
composition resulting from phase separation as WDs crystallize. To
this end, we consider phase diagrams suitable for $^{16}$O and
$^{20}$Ne plasmas \citep[][]{2010PhRvE..81c6107M}. We show that the
changes in the chemical profiles resulting from phase-separation
processes leave strong signatures in the theoretical pulsational
spectrum and must be taken into account in realistic computations of
the pulsational properties of ultra-massive WDs. Finally, element
diffusion was included for all model sequences, from the beginning of
the  WD cooling track. Element diffusion smoothes the inner chemical
profiles, thus altering the run of the Brunt-V\"ais\"al\"a frequency,
and hence the period spectrum and mode-trapping properties. The
pulsational study presented in this work is based on evolving models
of ultra-massive WDs recently computed in Camisassa et al. (2018,
submitted). These evolutionary sequences that incorporate all these
improvements were computed from the very beginning of the cooling
track down to very low surface luminosities.  The paper is organized
as follows. A summary of the numerical codes and the treatment of
crystallization we employed is provided in Sect. \ref{numerical}. In
Sect. \ref{evolutionary} we  briefly describe the evolutionary models
and the core chemical redistribution that is due to phase separation
during crystallization.  In Sect. \ref{pulsation} we present a
detailed description of the pulsation computations. In particular, in
Sect. \ref{IMPACT} we assess the impact of crystallization on the
pulsation spectrum of our ONe-core WD models, and in
Sect. \ref{CO-ONE-COMPARISON} we compare the pulsation properties of
our ONe-core WD models with those of CO-core WDs for a fixed mass
value. Finally, in Sect. \ref{conclusions} we summarize the main
findings of this work.

\section{Numerical tools}  
\label{numerical}  

\subsection{Evolutionary code}
\label{LPCODE}

The DA WD evolutionary models developed in this work were computed
with the {\tt LPCODE} evolutionary code \cite[see][for detailed
  physical description]{2005A&A...435..631A,2010ApJ...717..897A,
  2010ApJ...717..183R,2012MNRAS.420.1462R,2016A&A...588A..25M}. This
numerical tool has been employed to study various aspects of the
evolution of low-mass stars \citep{2011A&A...533A.139W,
  2013A&A...557A..19A,2015A&A...576A...9A}, the formation of
horizontal branch stars \citep{2008A&A...491..253M}, extremely
low-mass WDs \citep{2013A&A...557A..19A}, AGB, and post-AGB evolution
\citep{2016A&A...588A..25M}, among others. More recently, the code has
been used to assess the impact of the uncertainties in progenitor
evolution on the pulsation inferences of ZZ Ceti stars
\citep{2017A&A...599A..21D,2018A&A...613A..46D}. Next we describe the
main input physics of the code that are relevant for computing the
ultra-massive WD models: $i)$ convection is treated according to the
mixing length formulation \citep[ML2, ][]{1990ApJS...72..335T}; $ii)$
radiative and conductive opacities are those from OPAL
\citep{1996ApJ...464..943I} and \citet{2007ApJ...661.1094C},
respectively; $iii)$ the molecular radiative opacities for the
low-temperature regime are those from \citet{2005ApJ...623..585F};
$iv)$ the equation of state for the low-density regime is taken from
\citet{1979A&A....72..134M} and that for the high-density regime from
\citet{1994ApJ...434..641S}, which accounts for both the solid and
liquid phases; $v)$ element diffusion, including gravitational
settling and chemical and thermal diffusion, is considered; and $vi)$
energy release from crystallization (latent heat and gravitational
energy associated with ONe phase separation) are included by
considering a phase diagram that is suitable for the dense ONe
interiors of ultra-massive WD \citep{2010PhRvE..81c6107M}. To our
knowledge, this is the first pulsational analysis of ultra-massive WD
models that includes the phase-separation processes in ONe cores.

\subsection{Treatment of crystallization and phase separation}

Cool WD stars are expected to crystallize as a result of strong
Coulomb interactions in their very dense interior
\citep{1968ApJ...151..227V}.  Crystallization sets in when the energy
of the Coulomb interaction between neighboring ions is much higher
than their thermal energy.  This occurs when the ion coupling constant
$\Gamma \equiv \langle Z^{5/3} \rangle e^2/a_{\rm e} k_{\rm B} T$ is
larger than a certain value, which depends on the adopted phase
diagram. Here $a_{\rm e}$ is the inter-electronic distance, $\langle
Z^{5/3} \rangle$ is an average (by number) over the ion charges, and
$k_{\rm B}$ is Boltzmann's constant.  The other symbols have their
usual meaning.  The occurrence of crystallization leads to two
additional energy sources: the release of latent heat, and the release
of gravitational energy associated with changes in the chemical
composition profile induced by crystallization
\citep{1988A&A...193..141G,1988Natur.333..642G,2009ApJ...693L...6W}.
In our study, these two additional energy sources are included
self-consistently and locally coupled to the full set of equations of
stellar evolution. In particular,  the luminosity equation is
appropriately modified to account for both the local contribution of
energy released from the core chemical redistribution and the latent
heat.  At each time step, the crystallization temperature and the
change in chemical profile resulting from phase separation are
computed using the appropriate phase diagram. In particular, the
oxygen-enhanced convectively unstable liquid layers overlying the
crystallizing core are assumed to be instantaneously mixed, a
reasonable assumption considering the long evolutionary timescales of
WDs.  The chemical redistribution due to phase separation and the
associated release of energy  have been considered following the
procedure described in \citet{2010ApJ...719..612A}, appropriately
modified by Camisassa et al. (2018) for ONe plasmas.
To assess the enhancement of $^{20}$Ne in the crystallized core, we
used the azeotropic-type phase diagram of
\citet{2010PhRvE..81c6107M}. It is worth mentioning that this is a
two-component phase diagram, even though our WD models contain trace
amounts of other elements in the crystallized regions.
After computing the chemical composition of both the solid and the
liquid phases, we evaluated the net energy released in the process as
in \citet{2010ApJ...719..612A}; see also \citet{1997ApJ...485..308I}. 

\subsection{Pulsation code}
\label{LP-PUL}

We computed nonradial $g$-mode pulsations of our complete set of
ultra-massive ONe-core DA WD models using the adiabatic version of the
{\tt LP-PUL} pulsation code described in
\citet{2006A&A...454..863C}. The pulsation code is based on the
general Newton-Raphson technique that solves the full fourth-order set
of equations and boundary conditions governing linear, spheroidal,
adiabatic, nonradial stellar pulsations following the dimensionless
formulation of \citet{1971AcA....21..289D}.  We did not consider
torsional modes, since these modes are characterized by very short
periods  \citep[up to 20 s; see][]{1999ApJ...526..976M} and are not
observed in ZZ Ceti stars. To account for the effects of
crystallization on the pulsation spectrum of $g$-modes, we adopted the
``hard sphere'' boundary conditions, which assume that the amplitude
of the eigenfunctions of $g$-modes is drastically reduced below the
solid/liquid interface because of the non-shear modulus of the solid,
as compared with the amplitude in the fluid region
\citep[see][]{1999ApJ...526..976M}. In our code, the inner boundary
condition is not the stellar center, but instead the mesh-point
corresponding to the crystallization front moving toward the surface
\citep[see][]{2004A&A...427..923C,2005A&A...429..277C,2013ApJ...779...58R}.
Specifically, the hard-sphere boundary condition at the radial shell
corresponding to the outward-moving crystallization front
$(r_{\rm c}= r(M_{\rm  c}))$ reads

\begin{equation} 
  y_1= 0,\ \ \ y_2= {\rm arbitrary}, \ \ \ \ell y_3-y_4= 0. 
\label{hard}
\end{equation}

Here $y_1$ and $y_2$ represent  the  radial  and  horizontal
displacements, respectively, and $y_3$ and $y_4$ are the Eulerian
perturbation of the gravitational potential and its derivative.  The
last condition is the same as for the normal case in which the core is
in a fluid state and the boundary condition is applied at the stellar
center;  see Appendix B of \citet{1999ApJ...526..976M}.

The asymptotic period spacing is computed as in
\citet{1990ApJS...72..335T}.  For  $g$-modes  with  high   radial
order $k$  (long  periods),  the separation  of consecutive  periods
($|\Delta  k|= 1$)  becomes nearly constant  at a  value  given  by
the  asymptotic  theory of  nonradial stellar  pulsations.
Specifically,  the  asymptotic  period  spacing is given by

\begin{equation} 
\Delta \Pi_{\ell}^{\rm a}= \Pi_0 / \sqrt{\ell(\ell+1)},  
\label{aps}
\end{equation}

\noindent where

\begin{equation}
%\label{asympeq}
\Pi_0= 2 \pi^2 \left[\int_{r_1}^{r_2} \frac{N}{r} dr\right]^{-1}.
\label{p0}
\end{equation}

\noindent The squared Brunt-V\"ais\"al\"a frequency ($N$, one of the
critical frequencies of nonradial stellar pulsations) is computed as

\begin{equation}
%\label{bvf}
N^2= \frac{g^2 \rho}{P}\frac{\chi_{\rm T}}{\chi_{\rho}}
\left[\nabla_{\rm ad}- \nabla + B\right],
\label{bv}
\end{equation}

\noindent where the compressibilities are defined as
\citep{1990ApJS...72..335T,1991ApJ...367..601B}

\begin{equation}
\chi_{\rho}= \left(\frac{\partial ln P}{\partial ln \rho}\right)_{{\rm T}, \{\rm X_i\}}\ \ \
\chi_{\rm T}= \left(\frac{\partial ln P}{\partial ln T}\right)_{\rho, \{\rm X_i\}}.
\end{equation}

\noindent The Ledoux term $B$ is computed as 

\begin{equation}
%\label{B}
B= -\frac{1}{\chi_{\rm T}} \sum_1^{M-1} \chi_{\rm X_i} \frac{\partial ln
  X_i}{\partial ln P},
\label{BLedoux}
\end{equation}

\noindent where
\begin{equation}
\chi_{\rm X_i}= \left(\frac{\partial ln P}{\partial ln X_i}\right)_{\rho, {\rm T},
  \{\rm X_{j \neq i}\}}.
\end{equation}

The computation of the Ledoux term $B$ has been generalized in order
to include the effects  of having  multiple chemical species ($^{1}$H,
$^{4}$He, $^{12}$C, $^{16}$O, $^{20}$Ne, $^{23}$Na, $^{24}$Mg) that
vary  in  abundance. 

When a fraction of the WD core is crystallized, the lower limit of the
integral in Eq. (\ref{p0}) coincides with the radius of the
crystallization front ($r_1= r_{\rm c}$), which moves outward as the
star cools down and the fraction of crystallized mass
increases. Hence, the integral in Eq.( \ref{p0}) decreases, leading to
an increase in the asymptotic period spacing (Eq. \ref{aps}) and also
in the periods themselves.

\section{Evolutionary models}  
\label{evolutionary}  

We computed the evolution and pulsation properties of four
ultra-massive WD sequences with stellar masses $M_{\star}= 1.10, 1.16,
1.22$, and $1.29~ M_{\sun}$ resulting from the complete evolution of
the progenitor stars through the S-AGB phase.  The core  and
inter-shell chemical profiles of our models at the start of the  WD
cooling phase were obtained from \citet{2010A&A...512A..10S}. The
cores are composed mostly of $^{16}$O and $^{20}$Ne and smaller
amounts of $^{12}$C, $^{23}$Na, and $^{24}$Mg.  Since element
diffusion and gravitational settling operate throughout the WD
evolution, our models develop pure hydrogen envelopes. In Table
\ref{tabla1} we show the $^{1}$H and $^{4}$He mass content for each
sequence, as well as the $T_{\rm eff}$ and $\log g$ values
corresponding to the onset of crystallization, and the fraction of
crystallized mass at the boundaries of the ZZ Ceti instability strip.
The He content of our WD sequences is  given by the evolutionary
history of progenitor star, but instead, the H content  ($M_{\rm
  H}\sim 10^{-6}M_{\star}$) has been set by imposing that the further
evolution  does not lead to H thermonuclear flashes on the WD cooling
track. The higher mass model sequences start to crystallize  at effective
temperatures well above the instability strip, thus harboring a core
that is almost completely crystallized at the time the sequences
reach the ZZ Ceti stage. As consequence of element diffusion, the
chemical profiles at low $T_{\rm eff}$ are far smoother than those at
the beginning of the WD stage (see Figs. 8 and 9 of Camissasa et
al. 2018). In particular, the He buffer, which is located above the
ONe core, is strongly eroded, therefore developing a region rich in
$^4$He, $^{12}$C, $^{16}$O, and $^{20}$Ne.

\begin{table*}[]
\centering
%\resizebox{\columnwidth}{!}{%
\caption{H and He mass content of our ONe-core ultra-massive DA WD models, together with the effective temperature and surface gravity at the onset of crystallization, and the fraction of crystallized mass at the blue and red edges of the ZZ Ceti instability strip.}
\begin{tabular}{lcccccc}
\hline
  $M_{\star}/M_{\sun}$   &{\bf $M_{\rm H}/M_{\star}  $}&$M_{\rm He}/M_{\star}$    & $T_{\rm eff}^{\rm c}$&$\log g^{\rm c}$& $M_{\rm c}/M_{\star}$ & $M_{\rm c}/M_{\star}$ \\
  &($\times 10^{-6}$) & ($\times 10^{-5}$) & (K) & (cgs) & ($T_{\rm eff}= 12\,500$ K) & ($T_{\rm eff}= 10\,500$ K) \\
\hline
  1.098         &    1.5   & 29.6 & 19881  & 8.83     & 0.81                            &0.92                             \\
  1.159         &   1.5  & 15.7 & 23291  & 8.95     & 0.90                            &0.96                                 \\
  1.226         &   1.5  & 6.38 & 28425  & 9.12     & 0.96                            &0.98                                 \\
  1.292         &   1.5  & 1.66 & 37309  & 9.33     & 0.994                           &0.998                                \\
\hline
%\end{tabular}}
\end{tabular}
\label{tabla1}
\end{table*}

%\subsection{Crystallization and chemical redistribution caused by phase 
%separation}
%\label{PHASE_SEP}

\begin{figure} 
\includegraphics[clip,width=1.0\columnwidth]{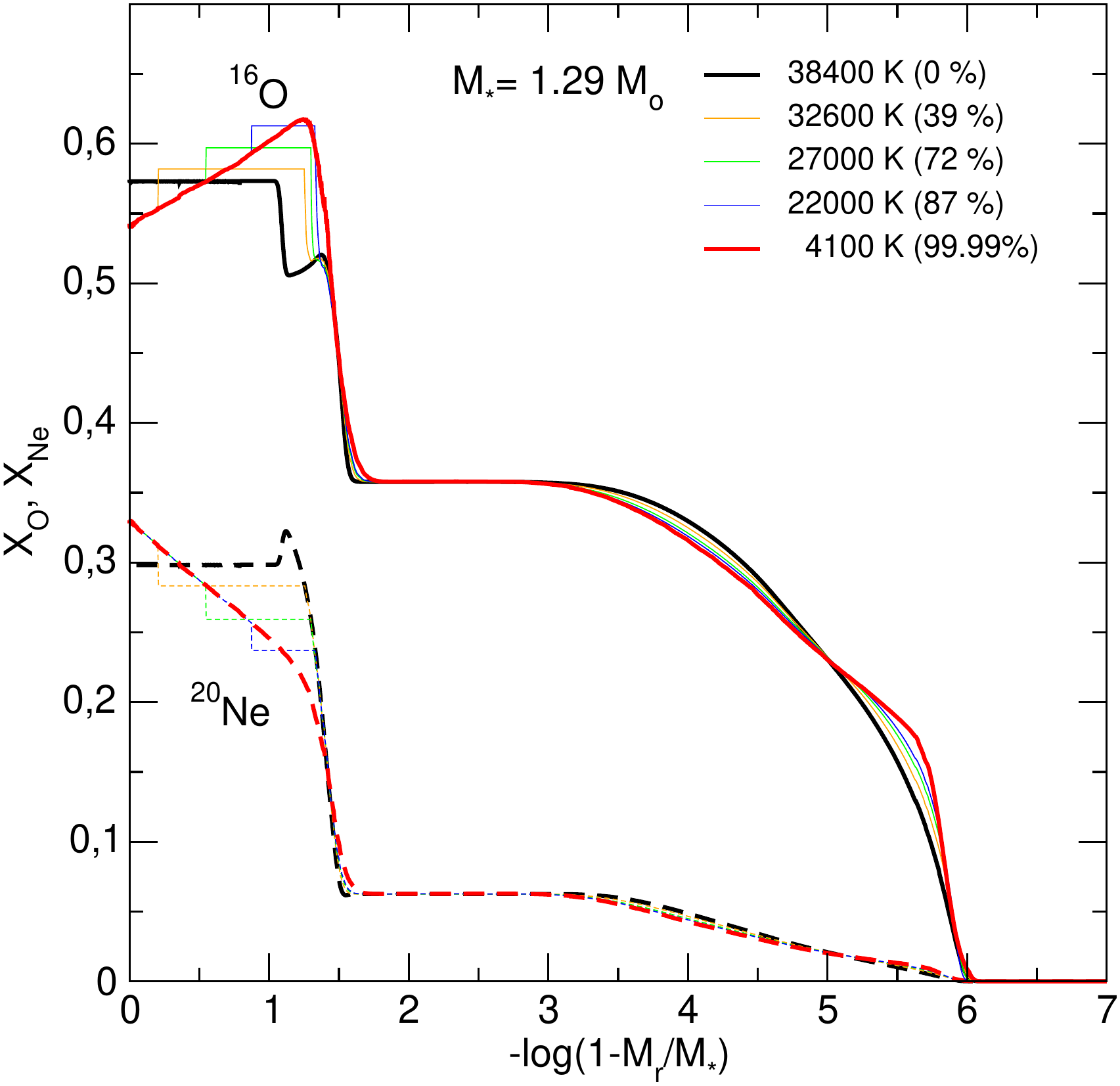} 
\caption{Internal chemical profiles of $^{16}$O (solid lines) and
  $^{20}$Ne (dashed lines) in terms of the fractional mass for the
  $1.29 M_{\sun}$ ONe-core WD sequence corresponding to various percentages of
  crystallization.  The thick black lines correspond to the profiles
  before crystallization, and the thin lines of
  different colors correspond to the chemical profiles resulting from
  chemical rehomogenization for different effective temperatures at
  increasing percentages of crystallized mass fraction, as indicated for some 
  selected cases.
  The thick red curves correspond to a model with $T_{\rm eff}= 4046$
  K and a percentage of $99.99$ \% of crystallized mass.}
\label{XONE-129} 
\end{figure} 

Theoretical evidence suggests that if the core of a WD is initially
composed of a mixture of $^{16}$O and $^{20}$Ne
\citep{2010PhRvE..81c6107M}, the crystallized region will have a
higher abundance of $^{20}$Ne than that in the original fluid
state. On the other hand, the liquid regions overlying the
crystallized $^{20}$Ne-enhanced layers will have a higher content of
$^{16}$O. This region will become Rayleigh-Taylor unstable, since
$^{16}$O is less dense than $^{20}$Ne. This instability leads to a
rehomogenization of the chemical profile in the fluid layers; as a
result of this, the chemical abundances after crystallization differ
substantially when compared with the initial ones. To compute the
chemical rehomogenization, we employed the same algorithm as in
\citet{1999ApJ...525..482M}, \citet{1997ApJ...486..413S}, and
\citet{2005A&A...429..277C}.  To derive the $^{20}$Ne enhancement when
a given layer crystallizes, we adopted the \cite{2010PhRvE..81c6107M}
azeotropic-type phase diagram for a $^{16}$O/$^{20}$Ne mixture. In our
scheme for mixing, we first considered a crystallized,
$^{20}$Ne-enhanced layer, and then we determined whether the innermost
fluid shell had a higher oxygen content than the overlying one.  If it
did, we mixed the two fluid layers and performed the same comparison
with the next layer farther out.  In this way, we scanned outward
through the fluid, and the process stopped when further mixing no
longer decreased the $^{20}$Ne content of the fluid between this point
and the crystallization boundary. When the crystallization front moved
outward due to cooling, the procedure was repeated.

In Fig. \ref{XONE-129} we show the $^{16}$O and $^{20}$Ne chemical
profiles in terms of the fractional mass for the $1.29 M_{\sun}$
ONe-core WD sequence corresponding to various degrees of
crystallization.  We did not include the trace elements --$^{12}$C,
$^{23}$Na, and $^{24}$Mg. Because we employed a two-component phase
diagram, these minoritary elements are considered inert during
crystallization.  The shape of the chemical profile is modified, even
in regions beyond the crystallization front. In particular, a 
  pronounced step is formed in the chemical profiles at the border of
the rehomogenized region. 

\section{Pulsation calculations}
\label{pulsation}

We computed adiabatic pulsation periods of $\ell= 1, 2$ $g$-modes in a
range of periods covering the period spectrum that is typically observed in ZZ
Ceti stars (70 s$\lesssim \Pi \lesssim 1500$ s). In our pulsation
computations, we considered three situations:

\begin{enumerate}
\item Crystallization does not take place (NC case).
\item Crystallization occurs, but only latent
  heat is released, and the core chemical profile remains unaltered (LH
  case).
\item Crystallization does take place, involving a release of latent
  heat and also a chemical redistribution as a result of phase separation, as
shown in Fig. \ref{XONE-129} (LH+PS case).
\end{enumerate}

\begin{figure} 
\includegraphics[clip,width=1.0\columnwidth]{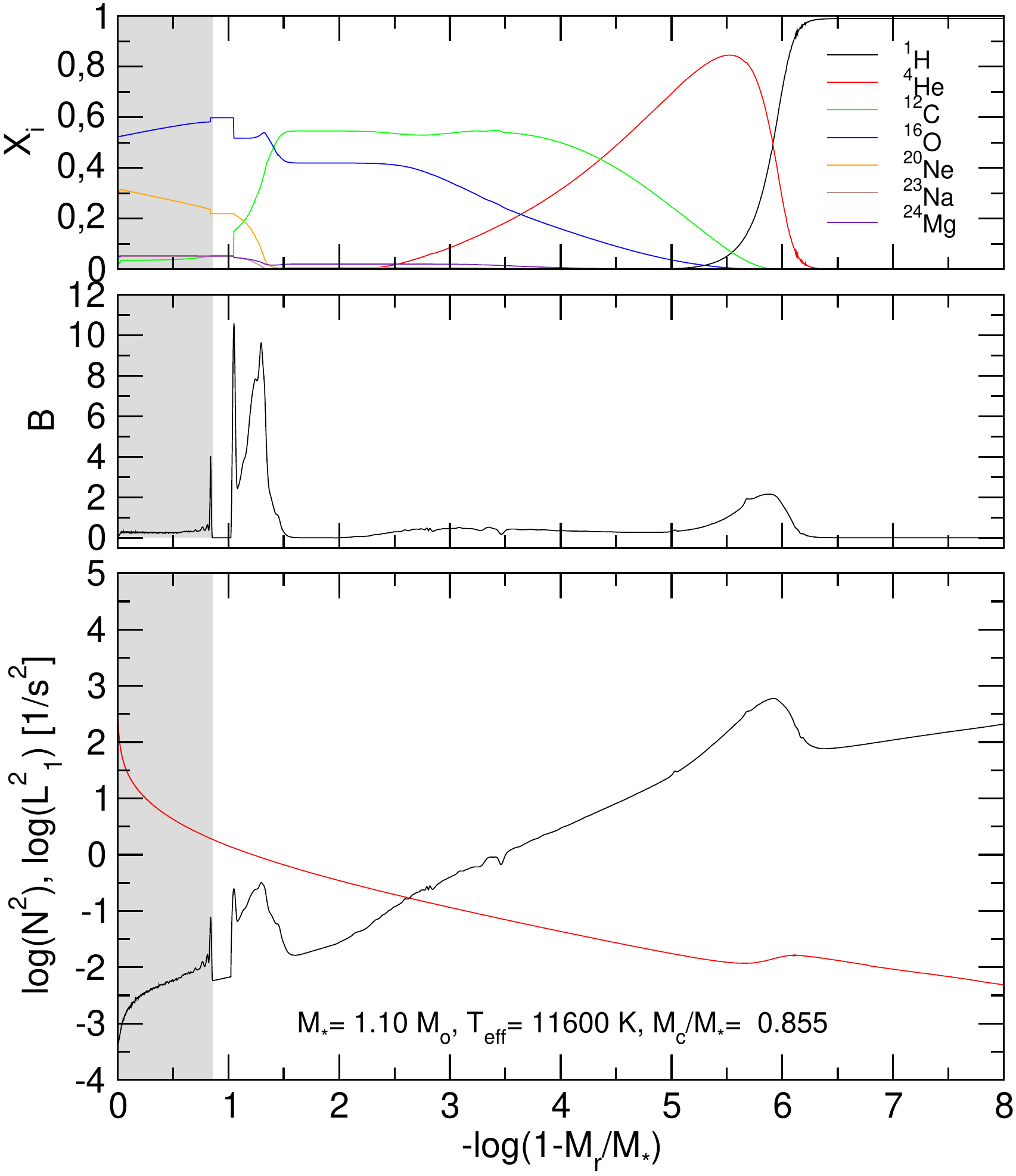} 
\caption{Abundances by mass of $^{1}$H, $^{4}$He, $^{12}$C,
  $^{16}$O, $^{20}$Ne, $^{23}$Na, and $^{24}$Mg as a function of the
  fractional mass (upper panel), the Ledoux term B (middle panel), and
  the logarithm of the squared Brunt-V\"ais\"al\"a and Lamb
  frequencies (lower panel), corresponding to an ONe-core WD model with
  $M_{\star}= 1.10 M_{\sun}$ and $T_{\rm eff} \sim 11\,600$ K. The model was computed taking into account
  latent heat release and chemical redistribution caused by phase separation
  during crystallization (LH+PS case). The
  gray area marks the domain of crystallization.  $M_{\rm
    c}/M_{\star}$ is the crystallized mass fraction of the
  model.}
\label{XBBVF-110} 
\end{figure} 

The NC case is considered merely to assess the period spectrum of
ultra-massive WDs when we neglect crystallization, that is, when the
ordinary boundary conditions at the stellar center are adopted to
solve the pulsation equations. The LH case is considered to show the
pulsation properties of ultra-massive WDs when a solid core is
included in computing the pulsational eigenspectra, that is, when we
use the hard-sphere boundary conditions to compute the eigenfunctions
of modes, but the core chemical profiles remain fixed as the WD
crystallizes. Finally, we consider the LH+PS case, which constitutes
the most physically plausible situation in which crystallization is
considered and the core chemical profiles are being modified
continuously during crystallization as a result of phase separation.

\begin{figure} 
\includegraphics[clip,width=1.0\columnwidth]{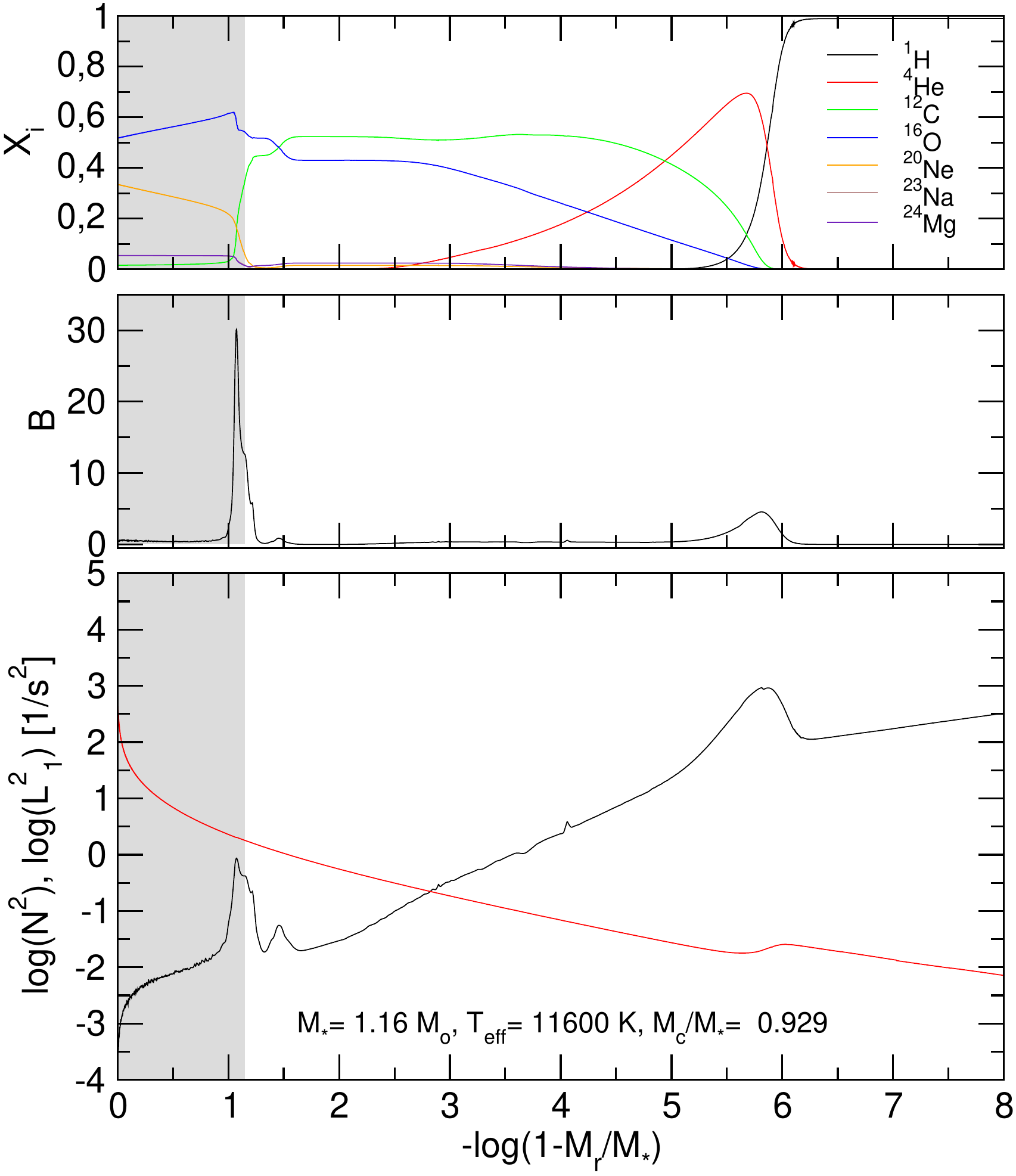}
\caption{Same as in Fig. \ref{XBBVF-110}, but for an ONe-core WD model with $M_{\star}=
  1.16 M_{\sun}$.}
\label{XBBVF-116} 
\end{figure} 

\begin{figure} 
\includegraphics[clip,width=1.0\columnwidth]{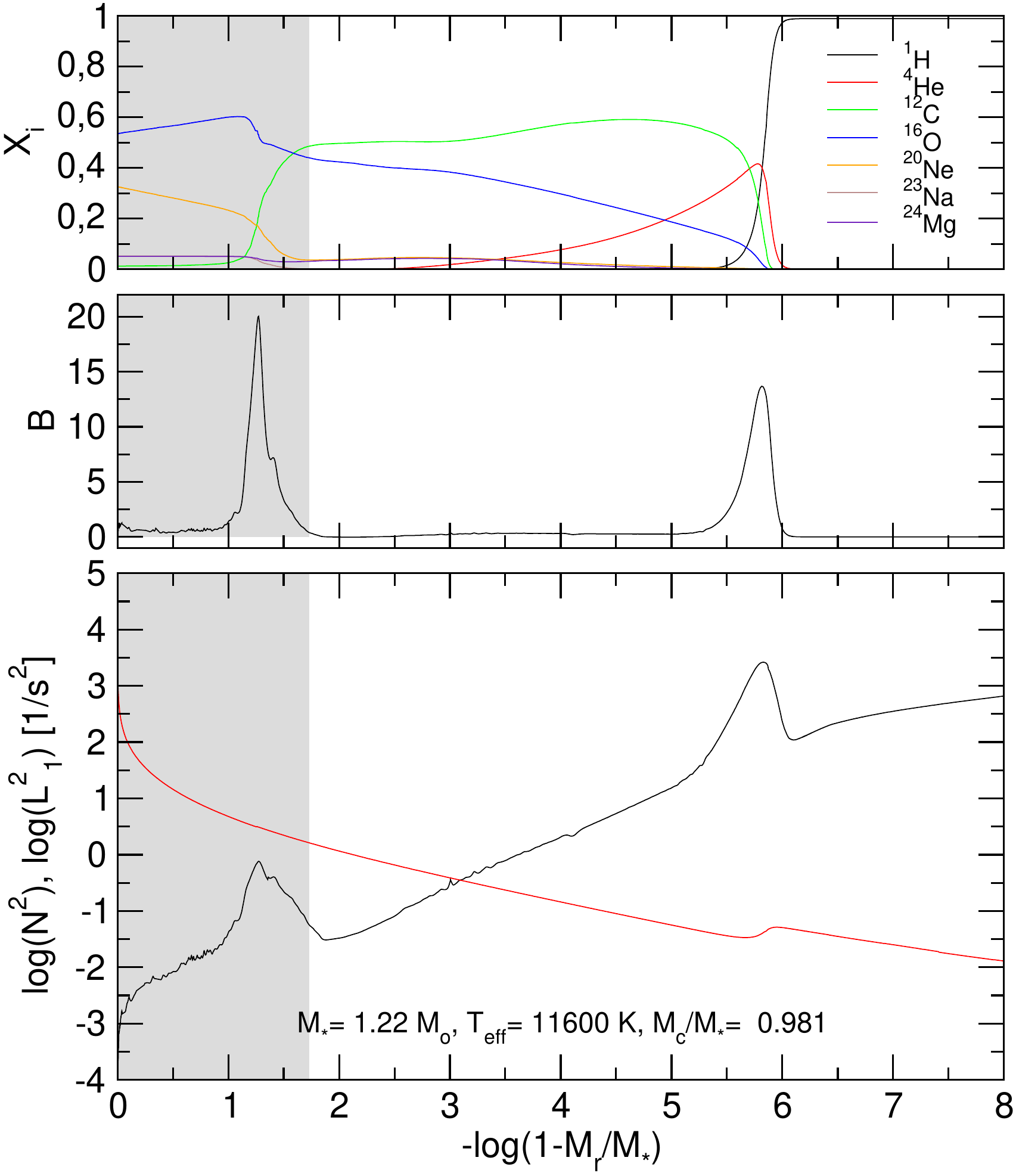} 
\caption{Same as in Fig. \ref{XBBVF-110}, but for an ONe-core WD model with $M_{\star}= 1.22 M_{\sun}$.}
\label{XBBVF-122} 
\end{figure}

\begin{figure} 
\includegraphics[clip,width=1.0\columnwidth]{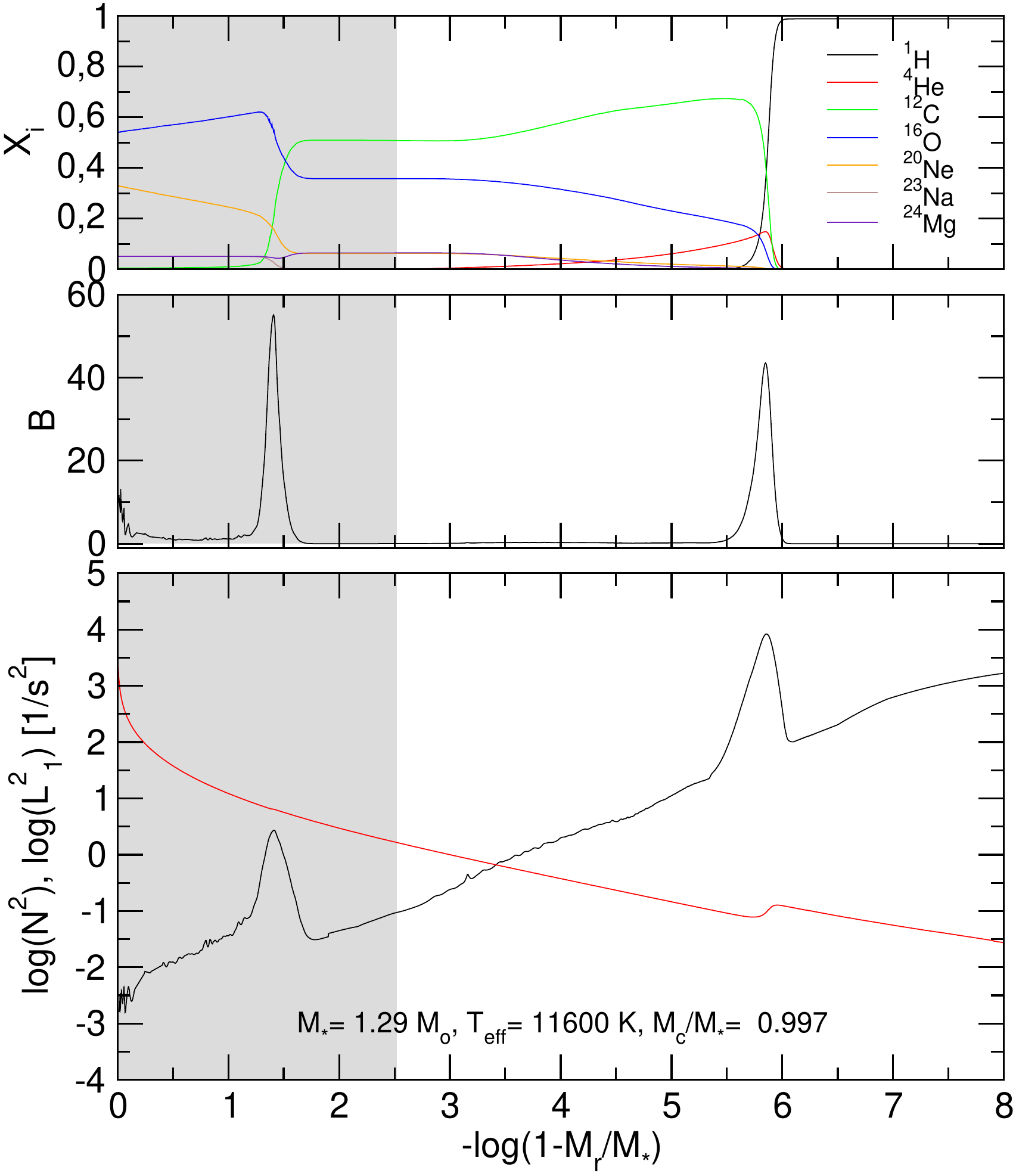} 
\caption{Same as in Fig. \ref{XBBVF-110}, but for an ONe-core WD model with $M_{\star}=
  1.29 M_{\sun}$.}
\label{XBBVF-129} 
\end{figure}

We show in Figs. \ref{XBBVF-110}, \ref{XBBVF-116}, \ref{XBBVF-122},
and \ref{XBBVF-129} the chemical abundances by mass of $^{1}$H,
$^{4}$He, $^{12}$C, $^{16}$O, $^{20}$Ne, $^{23}$Na, and $^{24}$Mg
(upper panel), the Ledoux term B (middle panel), and the logarithm of
the squared Brunt-V\"ais\"al\"a ($N$) and Lamb ($L_{\ell= 1}$)
frequencies (lower panel) in terms of the outer mass fraction
($-\log(1-M_r/M_{\star})$) for ONe-core WD models with
$M_{\star}/M_{\sun}= 1.10, 1.16, 1.22,$ and $1.29$, respectively, and
at an effective temperature in the middle of the ZZ Ceti instability
strip ($T_{\rm eff} \sim 11\,600$ K). The models were computed
taking into account crystallization with chemical redistribution that is due
to phase separation (LH+PS case).  The gray area marks the domain of
crystallization, and the fraction of the crystallized mass is
indicated for each model.

 For all the stellar masses considered, a notorious feature of the
  core chemical profiles is the redistribution of $^{16}$O and
  $^{20}$Ne at the solid core that is due to phase separation, and the
  consequent chemical rehomogeneization at layers beyond the
  crystallization front, as can be seen in Fig. \ref{XONE-129} for the
  case of the sequence with $M_{\star}= 1.29 M_{\odot}$. The range of
  effective temperatures at which phase separation takes place depends
  on the stellar mass value.  Specifically, phase separation during
  crystallization happens at higher $T_{\rm eff}$ for larger
  $M_{\star}$. In the case of the sequence with $M_{\star}= 1.10
  M_{\odot}$, crystallization and phase separation occurs for
  effective temperatures characteristic of  the ZZ Ceti instability
  strip.  Fig. \ref{XBBVF-110} illustrates the situation for a model
  of this sequence at $T_{\rm eff} \sim 11\,600$ K. An abrupt step is
  visible in the chemical profiles at the border of the
  rehomogeneization region,  at $-\log(1-M_r/M_{\star})\sim 1$.  This
  step is clearly reflected in the form of the Ledoux term $B$ as a
  very sharp peak, which is also visible in the shape of the
  Brunt-V\"ais\"al\"a frequency. Another notable feature in the
  chemical profiles is the triple interface of  $^{12}$C, $^{16}$O,
  and $^{20}$Ne, located at $1 \lesssim -\log(1-M_r/M_{\star})
  \lesssim 1.5$. This interface gives place to a bump in the
  Brunt-V\"ais\"al\"a frequency.  From $-\log(1-M_r/M_{\star})\sim 2$
  to $\sim 5$, there is a smooth transition region of $^{4}$He,
  $^{12}$C, and $^{16}$O, which leads to a very extended and almost
  negligible contribution to the Ledoux term $B$ and the
  Brunt-V\"ais\"al\"a frequency.  Finally, the remainder outstanding
  feature in the Ledoux term is the contribution of the
  $^{4}$He/$^{1}$H chemical transition region, which is translated
  into the shape of $N^2$ as a bump at $-\log(1-M_r/M_{\star}) \sim
  6$. The magnitude of this bump increases for the more massive
  models. We note that the innermost peak in B is due to the small
  step located exactly at the edge of the crystallized core. This also
  appears in the Brunt-V\"ais\"al\"a frequency as a small peak. In
  contrast with what occurs with the previously described features of
  $N^2$, the peak at the boundary of the crystallized core has no
  consequences for the pulsation spectrum because it is beyond the
  propagation zone of the $g$-modes.

In the case of the $1.16 M_{\odot}$ model (Fig. \ref{XBBVF-116}), at
$T_{\rm eff} \sim 11\, 600$ K the chemical redistribution of $^{16}$O
and $^{20}$Ne by phase separation has already finished, and the
resulting step at the border of the rehomogeneized region
($-\log(1-M_r/M_{\star})\sim 1$) is contained in the crystallized
region.  Consequently, the associated bump in the Brunt-V\"ais\"al\"a
frequency is irrelevant for the properties of $g$ modes, which do not
propagate in that region. In this way, the Brunt-V\"ais\"al\"a
frequency has only one bump which is able to inflict mode trapping
effects on  the modes, that is the bump at $-\log(1-M_r/M_{\star})\sim
6$ due to the $^{4}$He/$^{1}$H chemical interface.

The situation is similar for the models with $M_{\star}= 1.22
M_{\odot}$ (Fig. \ref{XBBVF-122}) and $M_{\star}= 1.29 M_{\odot}$
(Fig. \ref{XBBVF-129}). Indeed, in these cases the bump in the
Brunt-V\"ais\"al\"a frequency due to the triple transition region of
$^{12}$C, $^{16}$O, and $^{20}$Ne is excluded from the propagation
cavity of $g$ modes and the period spectrum is affected only by the
$^{4}$He/$^{1}$H chemical transition   region. 

 Finally, a detailed inspection of the central panels of
 Figs. \ref{XBBVF-110}, \ref{XBBVF-116}, \ref{XBBVF-122}, and
 \ref{XBBVF-129} in the region of the crystallized core reveals that
 the Ledoux term B has non-zero values as a result of the spatial
 variation of the abundances of $^{16}$O and $^{20}$Ne that are due to
 the chemical redistribution by phase separation. This small
 contribution of $B$ to the Brunt-V\"ais\"al\"a frequency, which is
 more notorious for the more massive models, is not relevant for their
 pulsation properties because the eigenfunctions of $g$-modes are
 excluded from the solid region.

\subsection{Impact of crystallization on the pulsation spectrum}
\label{IMPACT}

\begin{figure} 
\includegraphics[clip,width=1.0\columnwidth]{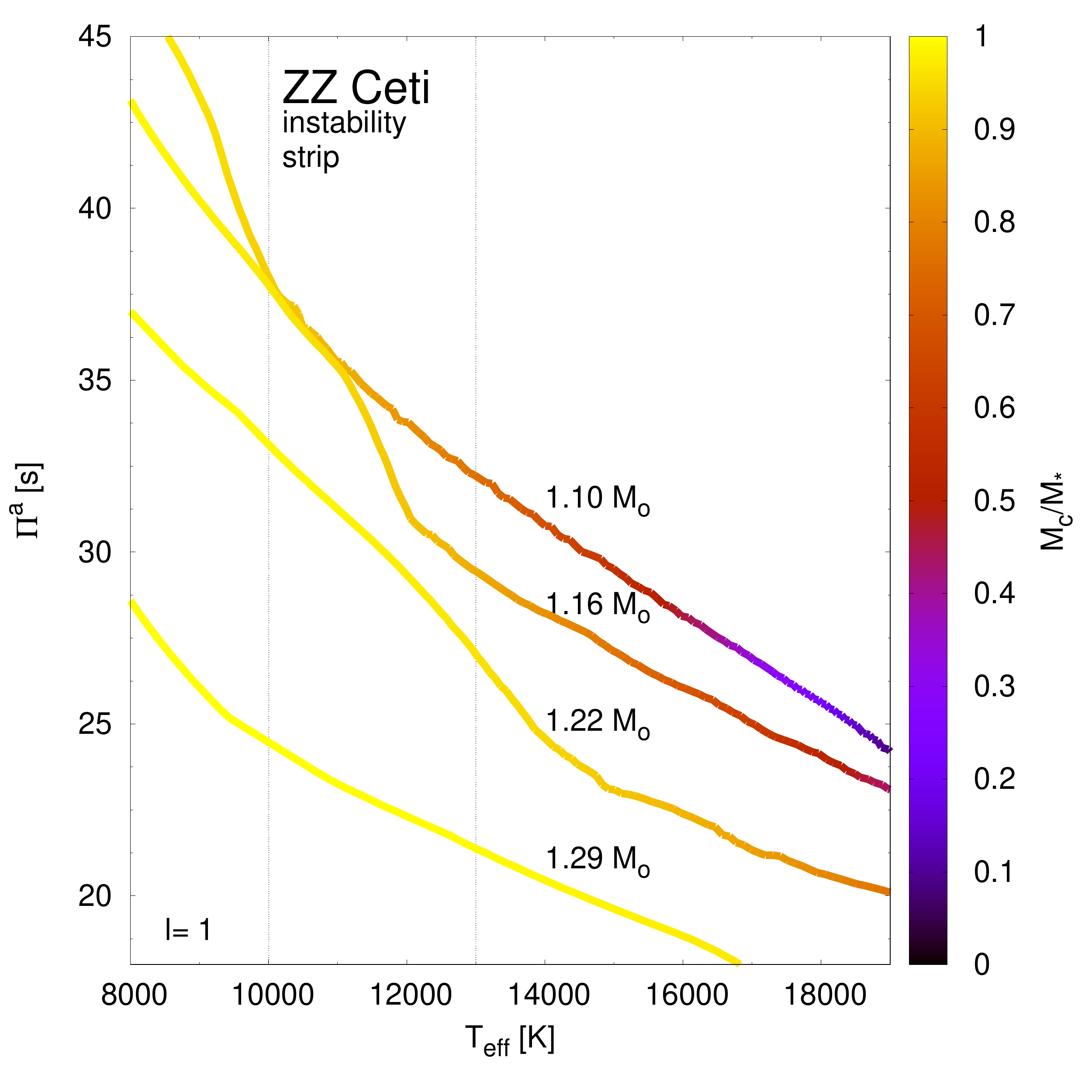}
\caption{Asymptotic period spacing, $\Delta \Pi^{a}$, for dipole
  modes ($\ell= 1$) as a function of the effective temperature for
  the ONe evolutionary cooling sequences with masses $1.10,
  1.16, 1.22,$ and $1.29 M_{\sun}$. Crystallization was computed taking
  into account latent heat release and chemical rehomogenization caused by
  phase separation. The palette of colors corresponds
  to the fraction of crystallized mass ($M_{\rm
    c}/M_{\star}$). Vertical dotted lines show the $T_{\rm eff}$
  interval of the ZZ Ceti instability strip.}
\label{APS4MASAS} 
\end{figure}

In this section we analyze the pulsation properties of our set
of ONe-core ultra-massive WD models. To do this, we compare the
impact of crystallization by comparing the results for the cases NC,
LH, and LH+PS.

\begin{figure}
\includegraphics[clip,width=1.0\columnwidth]{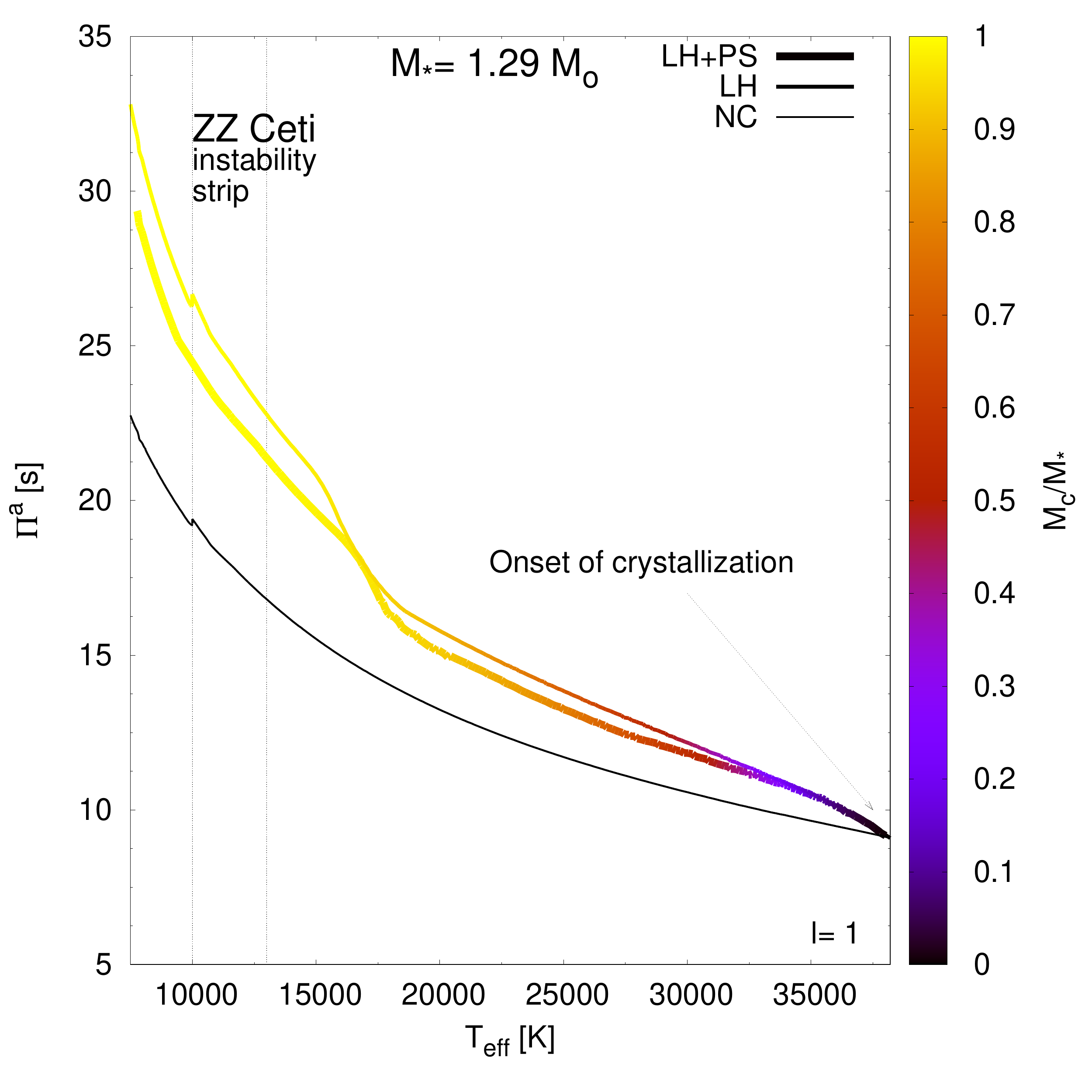} 
\caption{Dipole ($\ell= 1$) asymptotic period spacing as a
  function of the effective temperature for the ONe evolutionary
  cooling sequence with mass $M_{\star}= 1.29 M_{\sun}$. The thick
  curve corresponds to the case in which latent heat and chemical
  redistribution caused by phase separation have been taken into account
  during crystallization (LH+PS case), the intermediate-thickness
  curve displays the case in which chemical redistribution caused by
  phase separation has been neglected (LH case), and the thin curve is
  associated with the case in which crystallization has been
  neglected (NC case). The palette of colors corresponds to the fraction of
  crystallized mass ($M_{\rm c}/M_{\star}$). Vertical dotted lines
  show the $T_{\rm eff}$ interval of the ZZ Ceti instability strip.}
\label{APS129} 
\end{figure}

We begin by examining the asymptotic period spacing of our sequences,
computed according to Eqs. (\ref{aps}) and (\ref{p0}). In
Fig. \ref{APS4MASAS} we depict $\Delta \Pi_{\ell}^{a}$ for $\ell= 1$
modes in terms of $T_{\rm eff}$ for the ONe-core WD evolutionary
sequences with masses $1.10, 1.16, 1.22,$ and $1.29 M_{\sun}$. The
results correspond to the case in which crystallization was computed
taking into account latent heat release and chemical rehomogenization
caused by phase separation (LH+PS case).  In the case of the $1.22$
and $1.29 M_{\sun}$ sequences, the models have almost 99 \% of their
mass crystallized throughout the ZZ Ceti instability strip.  According
to Eqs. (\ref{aps}) and (\ref{p0}), the dependence of $\Delta
\Pi_{\ell}^{\rm a}$ on the Brunt-V\"ais\"al\"a frequency is such that
the asymptotic period spacing is larger when the mass and/or effective
temperature of the model is lower.  This trend is clearly visible in
Fig. \ref{APS4MASAS}.  The higher values of $\Delta \Pi_{\ell}^{\rm
  a}$ for lower $M_{\star}$ comes from the dependence $N \propto g$
(Eq. \ref{bv}), where $g$ is the local gravity ($g\propto
M_{\star}/R_{\star}^2$).  On the other hand, the higher values of
$\Delta \Pi_{\ell}^{\rm a}$ for lower $T_{\rm eff}$ result from the
dependence $N \propto \sqrt \chi_{T}$ (Eq. \ref{bv}), with $\chi_{T}
\rightarrow 0$ for increasing electronic degeneracy ($T \rightarrow
0$).

 A remarkable feature of $\Delta \Pi_{\ell}^{\rm a}$ is the abrupt
  change in the slope of the curves at certain effective
  temperatures. In the case of the $1.10 M_{\odot}$ sequence this
  happens at $T_{\rm eff} \sim 10\,000$ K, for the sequence with $1.16
  M_{\odot}$ this occurs at $T_{\rm eff} \sim 12\,000$ K, for $1.22
  M_{\odot}$ at $T_{\rm eff} \sim 14\,000$ K, and for $1.29 M_{\odot}$
  at $T_{\rm eff} \sim 17\,000$ K (not visible in
  Fig. \ref{APS4MASAS}). The change in the slope is due to the advance
  of the crystallization front.  At a given point of the evolution,
  this front surpasses the position of the triple interface of
  $^{12}$C, $^{16}$O, and $^{20}$Ne. This implies that the bump in the
  Brunt-V\"ais\"al\"a frequency associated to this interface ends up
  being inside the crystallized region and therefore excluded from the
  resonant cavity.  According to Eq. (\ref{p0}), the integral
  decreases abruptly when this happens, so that the asymptotic period
  spacing experiences a sudden growth (Eq. \ref{aps}).

In order to show the different expected behavior of the asymptotic
period spacing when we consider the NC, LH, and LH+PS treatments, we
display in Fig. \ref{APS129} $\Delta \Pi_{\ell= 1}^{\rm a}$ in terms
of the effective temperature for the ONe-core WD evolutionary sequence
with mass $M_{\star}= 1.29 M_{\sun}$. For this sequence, the onset of
crystallization occurs at $T_{\rm eff} \sim 37500$ K. For higher
effective temperatures, the three curves coincide. When the model
cools below this effective temperature, the curves of the asymptotic
period spacing corresponding to the LH and LH+PS cases begin to
separate from the curve associated with the NC case. The reason is, as
described before, that when crystallization begins, the lower limit of
the integral in Eq. (\ref{p0}) coincides with the radius of the
crystallization boundary, which continuously moves outward as the star
cools down. Thus, the fraction of crystallized mass increases and the
integral in Eq.(\ref{p0}) decreases, leading to an increase in $\Delta
\Pi_{\ell}^{\rm a}$ (Eq. \ref{aps}).  The asymptotic period spacing
for the LH and LH+PS cases is always longer than for the NC case, also
at the stages along the ZZ Ceti instability strip, where $\Delta
\Pi_{\ell}^{\rm a}$ is $\sim 4$ s shorter when crystallization is not
considered.   On the other hand, the value of $\Delta
  \Pi_{\ell}^{\rm a}$ for the LH case is slightly higher than for the
  LH+PS case, the larger difference (that amounts to $\sim 1.5$ s)
  being at the ZZ Ceti instability strip.
%This small
%difference is due to the presence of a peak in $N^2$ located at the
%end of the rehomegenized region in the LH+PS case, which is
%absent in the Brunt-V\"ais\"al\"a frequency for the LH case.
The case of the sequence with mass $M_{\star}= 1.29 M_{\sun}$ is
representative of the results for the sequences with masses $1.10,
1.16,$ and $1.22 M_{\sun}$, which are not shown for brevity.  Note the
abrupt increase of $\Delta \Pi_{\ell}^{\rm a}$ at effective
temperatures around $\sim 17\,000$ K for both the LH and LH+PS cases,
due to the reasons explained in detail in the previous paragraph.

\begin{figure}
\includegraphics[clip,width=1.0\columnwidth]{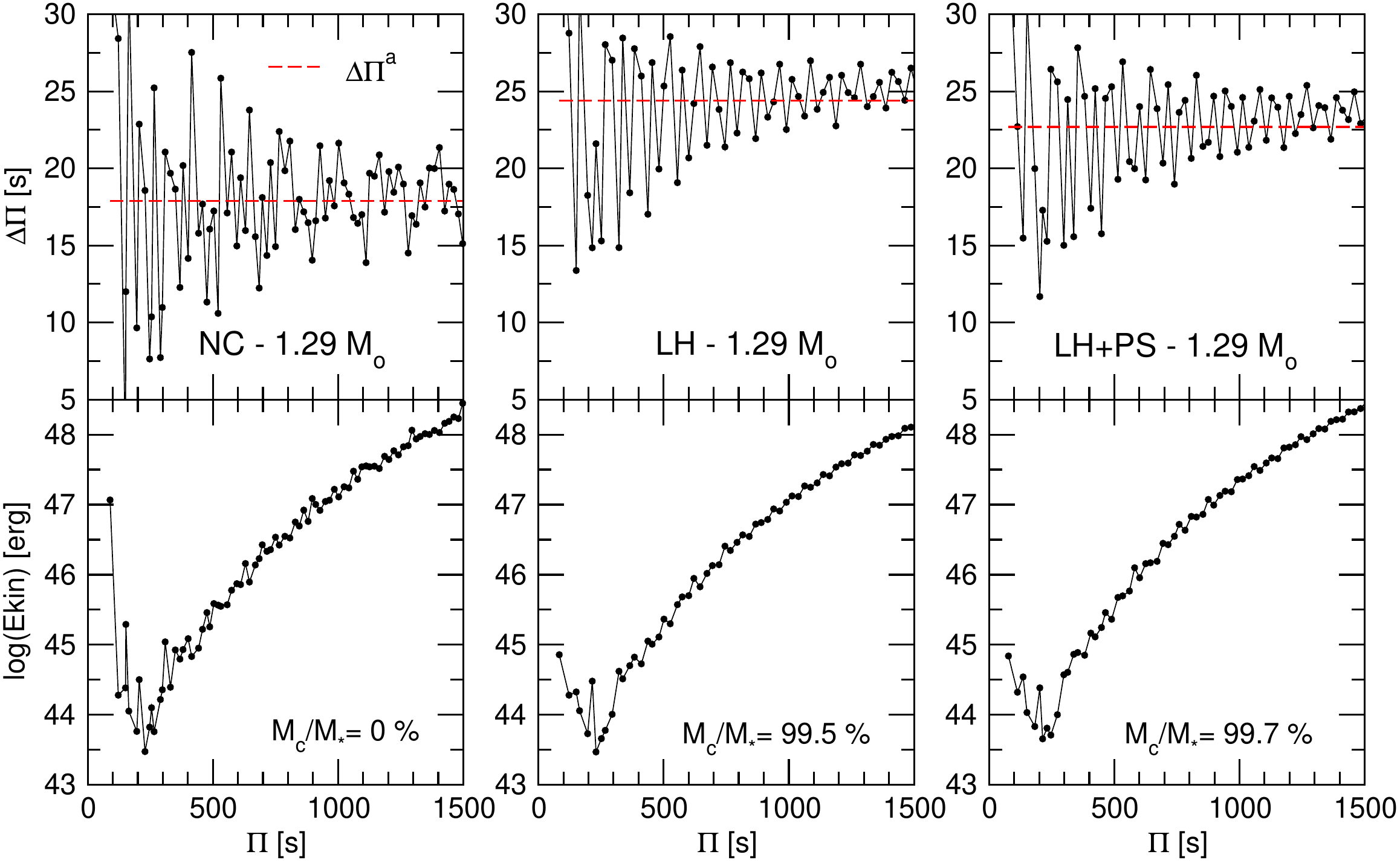} 
\caption{Forward period spacing ($\Delta \Pi$, upper panels)
  and logarithm of the kinetic energy ($E_{\rm kin}$, lower panels)
  in terms of the periods of $\ell= 1$  pulsation modes
  for a $1.29 M_{\sun}$ ONe-core WD model at $T_{\rm eff}  \sim 11\,600$ K.
  The left panels correspond to the case in which crystallization was
  not taken into account (NC case).
  The central panels show the same quantities for the situation in which
  crystallization was considered but
  phase separation was not (LH case). Finally, the right panels show the case
  in which both crystallization and phase separation were taken
  into account (LH+PS case). In the three cases, the percentage of
  the crystallized mass is indicated. In the upper panels, horizontal red dashed 
  lines correspond to the asymptotic period spacing.}
\label{DELPEKIN-1.29} 
\end{figure}

An important diagnostic tool for studying the mode-trapping
  properties in pulsating WDs is the $\Delta \Pi - \Pi$ diagram, in
  which the separation of periods with consecutive radial order $k$ is
  plotted in terms of the pulsation periods. A notable consequence of
  mode-trapping phenomena is that if we consider a fixed $T_{\rm eff}$
  value, the separation between consecutive periods has departures
  from the mean period spacing. This is clearly emphasized in the
  $\Delta \Pi - \Pi$ diagram shown in the right panels of
  Fig. \ref{DELPEKIN-1.29}, which show the forward period spacing
  ($\Delta \Pi_k \equiv \Pi_{k+1}-\Pi_k$.)  in terms of the periods of
  $\ell= 1$ pulsation modes for a $1.29 M_{\sun}$ ONe-core WD model at
  $T_{\rm eff} \sim 11\,600$ K for the LH+PC case (right), the LH case
  (middle panel) and the NC case (left panel). In all the cases,
  $\Delta \Pi_k$ exhibits maxima and minima typical of WD models
  harboring one or more chemical transition regions. The relationship
  between the values of the asymptotic periods spacing for the three
  cases follows the trend seen in Fig. \ref{APS4MASAS}. In particular,
  for the NC case, $\Delta \Pi_{\ell}^{\rm a}$ is markedly smaller
  than for the LH and LH+PS cases because the propagation cavity where
  the integral of the Eq. (\ref{p0}) is computed is quite greater when
  crystallization is neglected and $g$ modes can propagate to regions
  closer to the center of the star.

Regarding the mode-trapping patterns, the LH and LH+Ps cases (central
and right panels of Fig. \ref{DELPEKIN-1.29}) look very similar each
other. They  consist of alternating maxima and minima $\Delta \Pi_k$,
typical of WD models harboring a unique chemical interface. This is
expected, given that for a model at $T_{\rm eff} \sim 11\,600$ K, the
core chemical interfaces are inside the crystallized region, and the
only interface remaining at the propagation cavity is that of
$^{4}$He/$^{1}$H.  The mode-trapping pattern associated to the NC case
(left panel of Fig. \ref{DELPEKIN-1.29}), on the other hand, is a bit
more complex and this is due to the fact that there is more than one
chemical interface playing a role in the mode-trapping properties of
the model.

In the lower panels of Fig. \ref{DELPEKIN-1.29} we plot the logarithm
of the pulsation kinetic energy in terms of periods for $1.29
M_{\sun}$. We employed the usual normalization condition, that is,
that the value of the radial eigenfunction ($y_1=\xi_r/r$) is set to 1
at the surface of the model. The kinetic energy values plotted against
the periods exhibit virtually the same behavior  in the three cases
considered.

\begin{figure}
\includegraphics[clip,width=1.0\columnwidth]{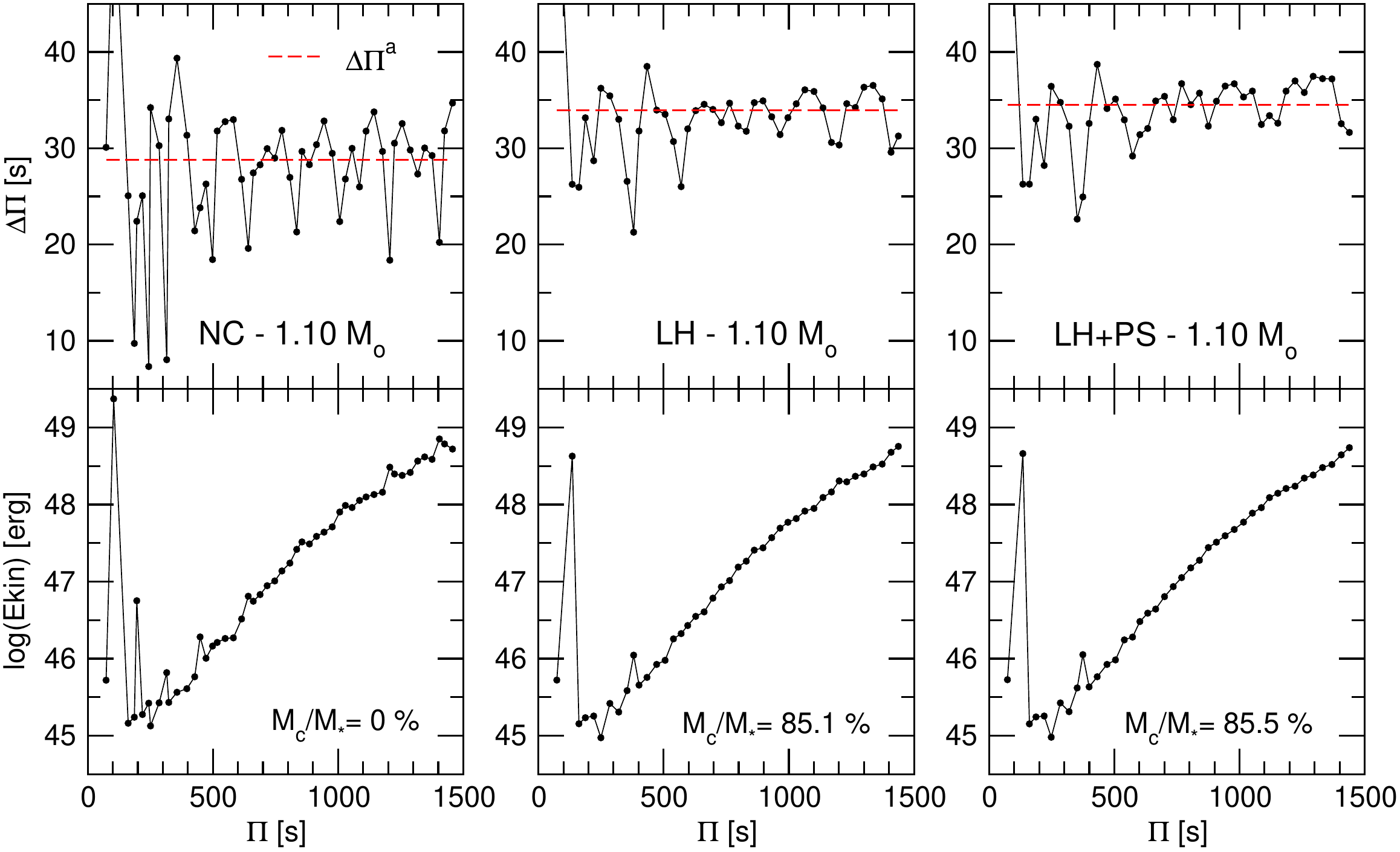} 
\caption{Same as Fig. \ref{DELPEKIN-1.29}, but for an ONe-core WD model with
  $M_{\star}= 1.10 M_{\sun}$.}
\label{DELPEKIN-1.10} 
\end{figure}

Fig. \ref{DELPEKIN-1.10} depicts the same as Fig. \ref{DELPEKIN-1.29},
but for WD models with $T_{\rm eff} \sim 11\,600$ K and $M_{\star}=
1.10 M_{\sun}$.  The mode-trapping behaviour associated to these
models is quite similar to that described for the $M_{\star}= 1.29
M_{\sun}$ models. In particular, the period-spacing patterns for LH
and LH+PS cases are virtually indistinguishable, whether we look at
the $\Delta \Pi_k$ values or if we look at the $\log(E_{\rm kin})$
values. On the other hand, the amplitudes of mode trapping are larger
for the NC case.

From the results shown in Figs. \ref{DELPEKIN-1.29} and
\ref{DELPEKIN-1.10}, we conclude that it would not be feasible to
distinguish the cases considered, that is NC, LH, or LH+PS from the
pulsation patterns of real ZZ Ceti stars, in particular from the
period spacing values.

\subsection{Comparison with CO-core WD models}
\label{CO-ONE-COMPARISON}

\begin{figure}
 \includegraphics[clip,width=1.0\columnwidth]{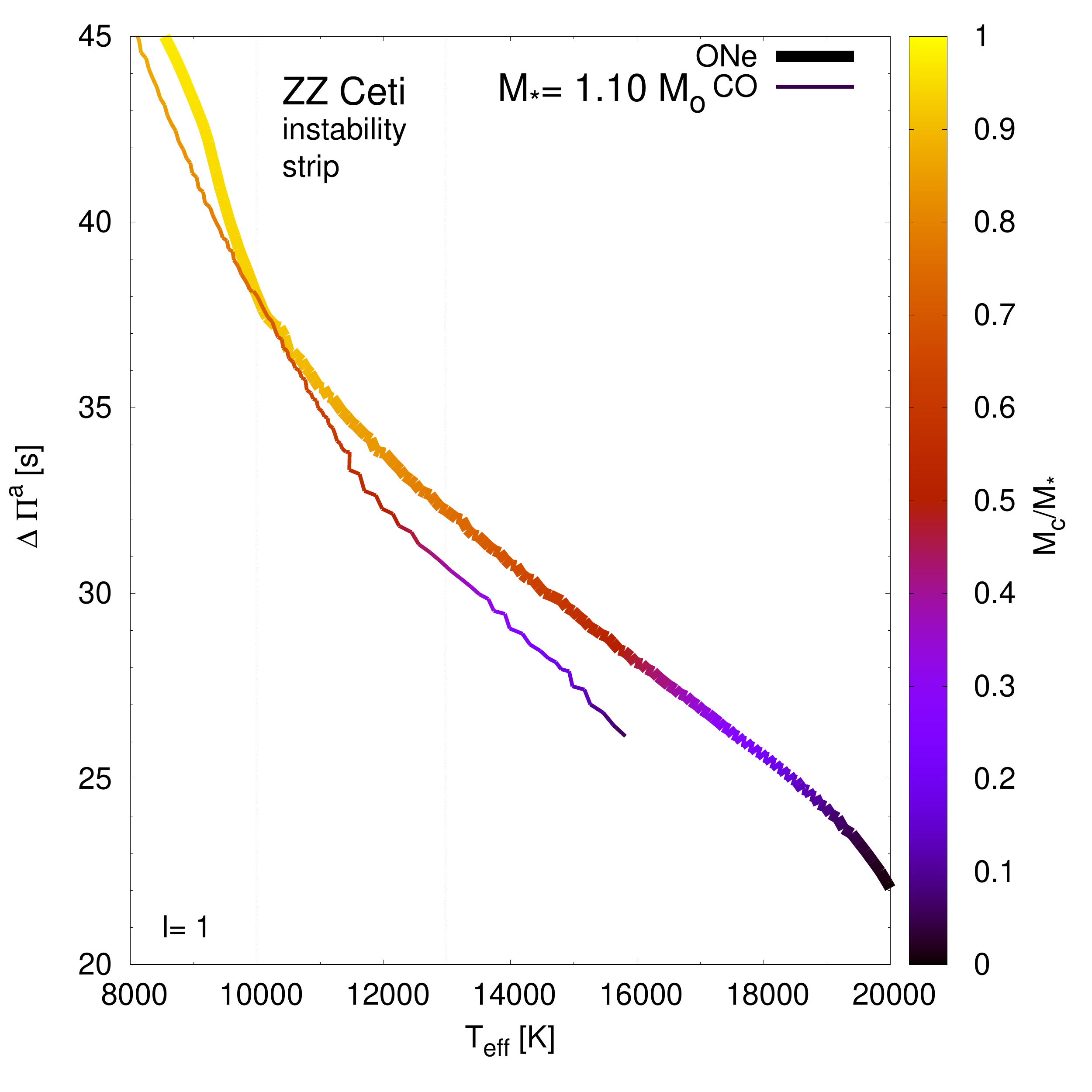} 
\caption{Dipole ($\ell= 1$) asymptotic period spacing
  as a function of the effective temperature for the ONe evolutionary cooling sequence
  (thick curve) and the CO evolutionary cooling sequence with mass $M_{\star}= 1.10 M_{\sun}$.
  Both sequences were  computed taking into account 
  latent heat release and chemical redistribution caused by phase separation during
  crystallization (LH+PS case).   Vertical dotted lines show the $T_{\rm eff}$ interval of the
  ZZ Ceti instability strip.}
\label{DELP-ASIN-ONE-CO} 
\end{figure}

The mass limit that separates the WDs harboring CO cores from those
with ONe cores is a matter of debate. It is therefore interesting to
compare the pulsation properties of WDs with CO cores and ONe cores
with the same stellar mass. The aim of this analysis is to explore the
possibility of using WD asteroseismology to distinguish between the
two types of objects. This has been precisely the goal of the study in
\citet{2004A&A...427..923C}, who compared the pulsation properties of
a $1.06 M_{\sun}$ WD model with CO core and a model with the same
mass, but a core made of O and Ne. Here, we re-examine the topic by
taking into account the chemical rehomogenization caused by phase
separation. This piece of physics was ignored in
\citet{2004A&A...427..923C} for the ONe-core WD model employed in that
work.

We computed an additional evolutionary sequence of WD models with
$M_{\star}= 1.10 M_{\sun}$ and a core made of $^{12}$C and $^{16}$O
for our analysis here, taking into account latent heat release and
chemical rehomogenization caused by phase separation during
crystallization (LH+PS treatment).  We employed the azeotropic phase
diagram for a mixture of C and O reported by
\citet{2010PhRvL.104w1101H}.  For this sequence, we computed the
adiabatic pulsation periods of $g$ -modes in the range of periods
observed in ZZ Ceti stars, as well as the asymptotic period spacing.
In Fig. \ref{DELP-ASIN-ONE-CO} we show the dipole ($\ell= 1$)
asymptotic period spacing as a function of $T_{\rm eff}$ for the ONe
evolutionary cooling sequence and the CO evolutionary cooling sequence
with stellar mass $M_{\star}= 1.10 M_{\sun}$. Both sequences were
computed taking into account latent heat release and chemical
redistribution due to phase separation during crystallization (LH+PS
case).  The figure shows that the asymptotic period spacing for
  both sequences is very similar, showing a modest difference of $\sim
  2$ s at the beginning of the ZZ Ceti instability strip (blue edge),
  and becoming virtually the same towards the red edge.  This is a
somewhat expected result given that the asymptotic period spacing of
$g$-modes in ZZ Ceti stars basically depends on the stellar mass, the
thickness of the H envelope, and the effective temperature, and these
three parameters are the same for both sequences. We note, however,
that by virtue of the different core chemical compositions, the degree
of crystallization for a given $T_{\rm eff}$ is different for both
types of WD models. We conclude that the asymptotic period spacing
 would not be a useful quantity for distinguishing ZZ Cetis
harboring CO cores from those with cores made of ONe.

\begin{figure} 
\includegraphics[clip,width=1.0\columnwidth]{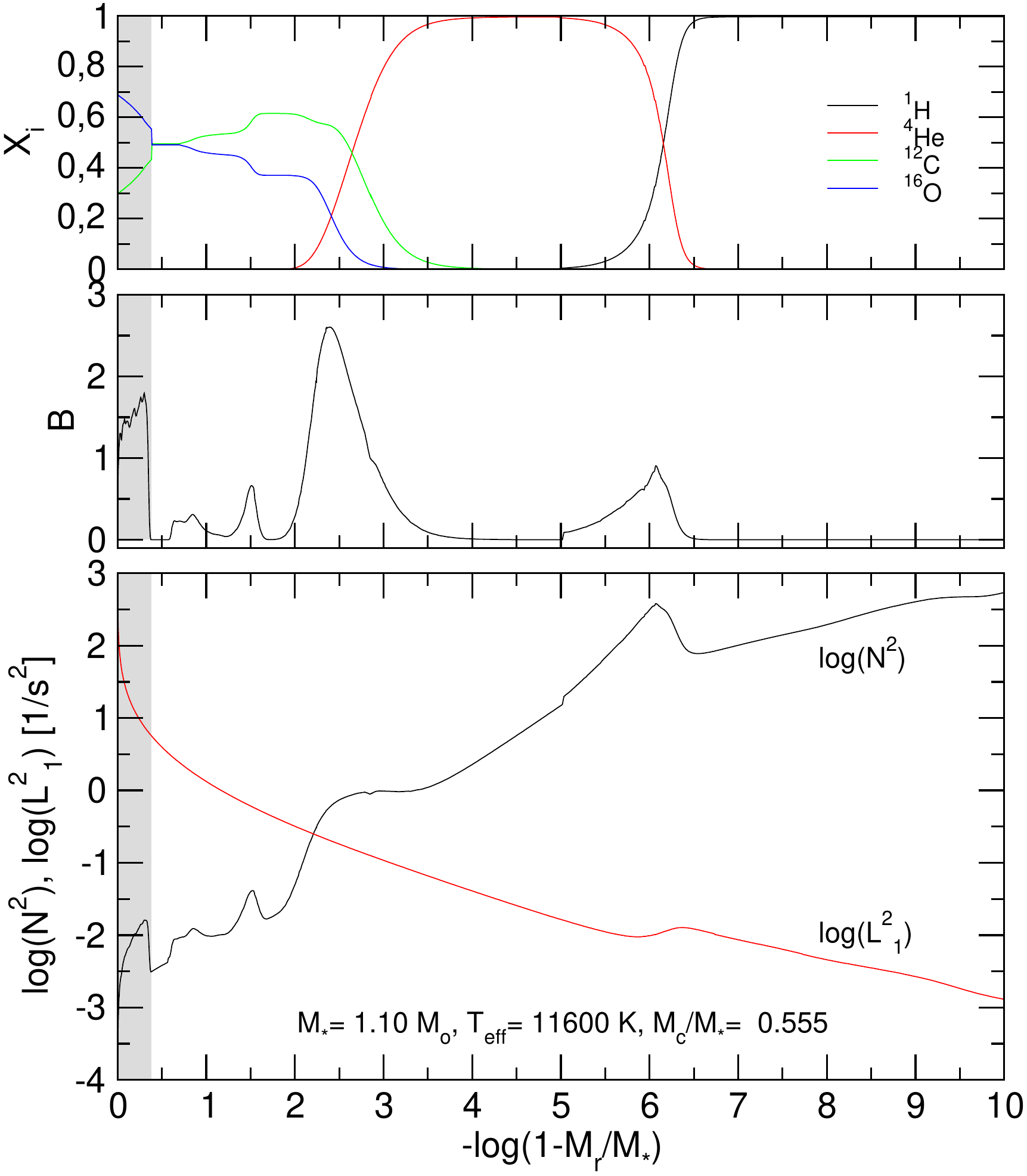} 
\caption{Abundances by mass of $^{1}$H, $^{4}$He, $^{12}$C, and $^{16}$O
  as a function of the fractional mass (upper panel), the Ledoux term B
  (middle panel), and the logarithm of the squared Brunt-V\"ais\"al\"a and Lamb
  frequencies (lower panel), corresponding to a CO-core WD model
  with $M_{\star}= 1.10 M_{\sun}$, $\log(M_{\rm H}/M_{\star})= -6$,  
  and $T_{\rm eff} \sim 11\,600$ K.
  Latent heat release and chemical redistribution caused by phase separation have been taken
  into account during crystallization (LH+PS case)
  The gray area marks the domain of crystallization.
  $M_{\rm c}/M_{\star}$ is the fraction of the crystallized mass of the model.}
\label{XBBVF-110-CO} 
\end{figure} 

\begin{figure}
\includegraphics[clip,width=1.0\columnwidth]{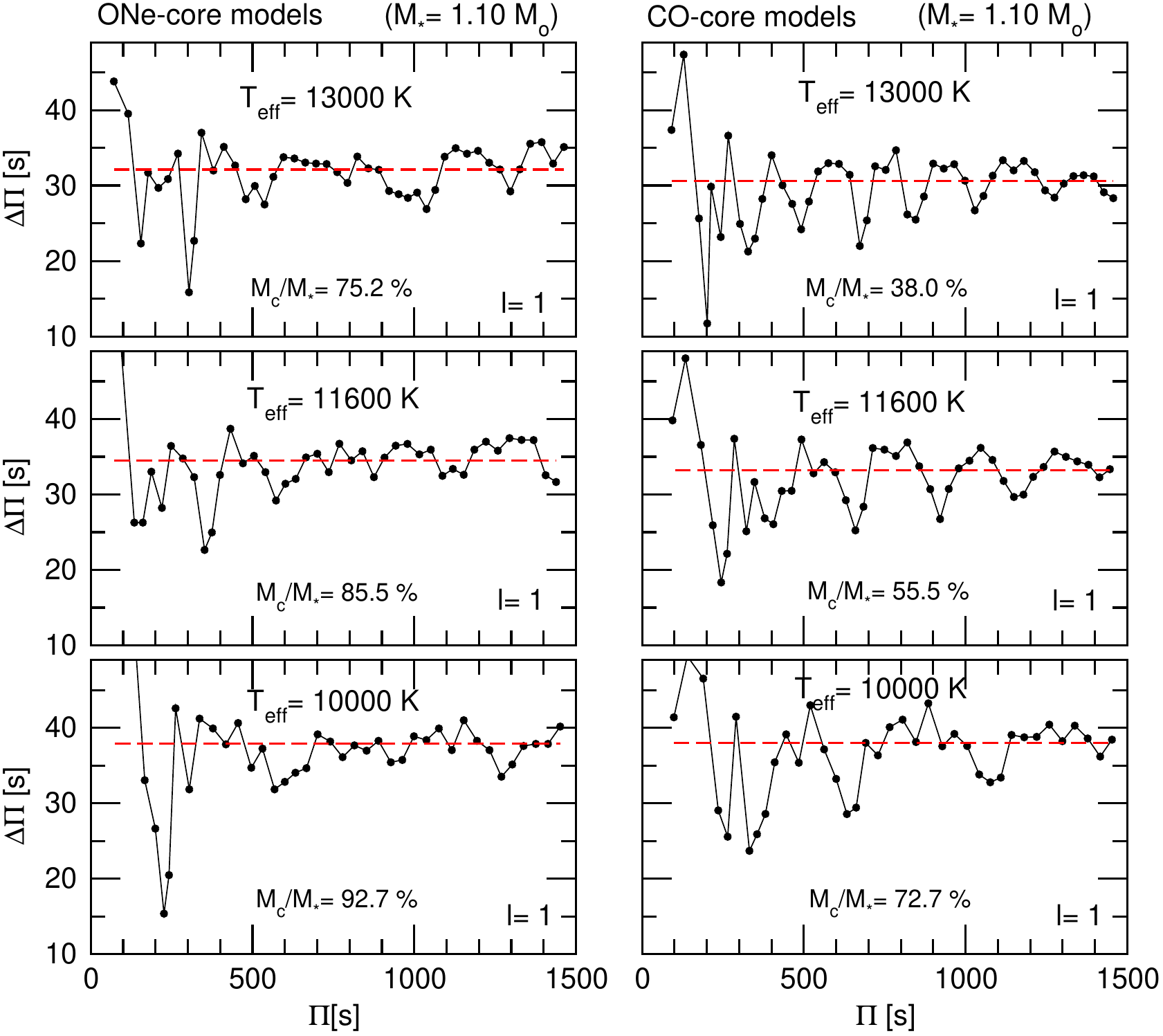} 
\caption{Forward period spacing ($\Delta \Pi$)
  in terms of the periods of $\ell= 1$  pulsation modes
  for a $1.10 M_{\sun}$ WD model at $T_{\rm eff}  \sim 13\,000$ K (upper panels),
  $\sim 11\,600$ K (middle panels), and $\sim 10\,000$ K (lower panels) with an
  ONe core (left panels) and a CO core (right panels). In both set of models,
  latent-heat release and chemical redistribution caused by phase separation have been taken
  into account during crystallization (LH+PS case). The percentage of
  the crystallized mass are indicated. The horizontal dashed red line is the asymptotic period
  spacing.}
\label{DELP-ONE-CO} 
\end{figure}

An alternative way to distinguish WDs with CO cores from those with
ONe cores for a fixed stellar mass is to examine the period-spacing
values in terms of the periods. This has been the approach employed
by C\'orsico et al. (2004). To this aim, we selected a $1.10 M_{\sun}$
CO-core template WD model at an effective temperature at the middle of
the ZZ Ceti instability strip ($T_{\rm eff} \sim 11\,600$ K). In
Fig. \ref{XBBVF-110-CO} we show the abundances by mass of $^{1}$H,
$^{4}$He, $^{12}$C, and $^{16}$O as a function of the fractional mass,
the Ledoux term B, and the logarithm of the squared
Brunt-V\"ais\"al\"a and Lamb frequencies of this template CO-core WD
model, with a crystallized mass of $\sim 56 \%$. In
Fig. \ref{DELP-ONE-CO} we plot the period spacing and the kinetic
oscillation energy in terms of the periods for the CO-core template
model and its counterpart ONe-core WD model (whose properties are
shown in Fig. \ref{XBBVF-110}). In the central panels we depict the
situation for $T_{\rm eff} \sim 11\,600$ K. Notably, $\Delta \Pi^{\rm
  a}$ is almost the same for both models, in agreement with the
discussion in the previous paragraph.  However, the period-spacing
  distribution is qualitatively different for both sets of
  models. Indeed, the pattern exhibited by the CO-core WD models is
  characterized by larger minima in $\Delta \Pi_k$ values for periods
  longer than $\sim 500$ s, as compared with the pattern shown by the
  ONe-core WD models.  Interestingly enough, this result is the
  opposed to that of \cite{2004A&A...427..923C}, who found that their
  ONe-core WD model showed stronger mode-trapping features than their
  CO-core WD model. This discrepancy in the results arises because
  \cite{2004A&A...427..923C}  considered chemical rehomogenization
  caused by phase separation in the  case of the WD model with CO
  core, but  not in the case of  the WD model with ONe core, whereas
  we here take this effect into account for both types of chemical
  compositions.  In addition, the chemical structure of our CO-core WD
  models largely differ from that of the models adopted by
  \cite{2004A&A...427..923C}. 

We conclude that the forward period spacing could
be employed to differenciate cores made of $^{12}$C and $^{16}$O, or
$^{16}$O and $^{20}$Ne, provided that a sufficient number of $g$ modes
with consecutive radial order are detected. Another plausible way to
determine whether an ultra-massive ZZ Ceti star has a CO core or a ONe
core is to carry out detailed asteroseismological studies involving
period-to-period comparisons of real stars employing two sets of
ultra-massive WD models, one of them characterized by CO-core WDs and
the other one with ONe-core WDs.  This analysis is beyond the scope of
this study and will be the focus of a future work.

\section{Summary and conclusions}
\label{conclusions}

We have assessed the adiabatic pulsation properties of ultra-massive
H-rich WDs with ONe cores on the  basis of full evolutionary models
that incorporate the most recent physical inputs governing the
progenitor and the WD evolution. This investigation constitutes a
substantial improvement over results reported by
\citet{2004A&A...427..923C} in several aspects. On one hand, the
chemical profiles of our WD models are consistent with the
predictions of the progenitor evolution through the S-AGB phase for
all the WD sequences with different stellar masses considered. On the
other hand, during WD evolution, we have taken into account for the
first time the changes in the core chemical composition resulting from
phase separation upon crystallization using phase diagrams suitable
for $^{16}$O and $^{20}$Ne plasmas. Finally, element diffusion was
included for all model sequences, from the beginning of the  WD
cooling track. Element diffusion smoothes the inner chemical profiles,
which strongly affects the run of the Brunt-V\"ais\"al\"a frequency,
and thus the period spectrum and mode trapping properties.

We assessed the pulsational properties of our models by computing
their dipole ($\ell= 1$) and quadrupole ($\ell= 2$) $g$-mode period
spectra for a wide range of effective temperatures, and in particular,
for the $T_{\rm eff}$ interval defining the ZZ Ceti instability strip.
Because our models are very massive ($1.10 \leq M_{\star}/M_{\sun}
\leq 1.29$), when they reach the ZZ Ceti instability strip, a very
large portion of their mass is in a crystalline phase. In order to
explore the impact of crystallization on their pulsation properties in
detail, we considered three cases: $(i)$ we neglected crystallization
in the equilibrium models and also in computing the pulsation modes,
$(ii)$ we considered crystallization with the ensuing release of
latent heat, and computed the pulsation modes adopting the hard-sphere
boundary conditions, and $(iii)$ we considered crystallization with
latent heat release and also took into account phase separation that
induces a chemical rehomogeneization in the liquid part surrounding
the crystallized core and the eigenmodes are computed using the
hard-sphere boundary conditions.  We did not find appreciable
  differences in the pulsation spectra of our WD models for the cases
  we analyzed. In particular, we find a very similar distribution of
  period spacings in the case in which the changes of the core
  chemical profiles that are due to phase separation are taken into
  account in comparison with the case in which it is ignored. In the
  case in which crystallization is ignored, the mode-trapping features
  in the periods-pacing diagram are slightly more pronnounced than in
  the cases in which crystallization is considered.

We also revisited the possibility of using asteroseismology to
distinguish ultra-massive DA WDs harboring ONe cores from those having
CO cores. In particular, we compared the pulsation properties
(specifically, periods and period spacings) of WD models characterized
by the same stellar mass ($1.10 M_{\sun}$), but in one case having a
core made of $^{16}$O and $^{20}$Ne and in another case a core
composed by $^{12}$C and $^{16}$O.  We did not find sizeable
  differences in the mean period spacing at the ZZ Ceti stage.
  However, we find appreciable differences in the period-spacing
  distribution,  in agreement with \cite{2004A&A...427..923C}.  In
  line with the  claim of \cite{2004A&A...427..923C}, we therefore
  conclude that period-spacing diagrams would allow to distinguishing
  the chemical composition of the cores of ultra-massive ZZ Ceti
  stars, provided that a sufficient number of $g$-mode periods with
  consecutive radial order are detected as to allow the construction
  of such period-spacing diagrams.  Additionally, in order to infer
  the core composition of  ultra-massive DA WDs, it will be profitable
  to carry out a detailed asteroseismic analysis using the individual
  periods observed in ultra-massive ZZ Ceti stars such as BPM 37093
  \citep {1992ApJ...390L..89K,2005A&A...432..219K}, GD 518
  \citep{2013ApJ...771L...2H}, and  SDSS J0840$+$5222
  \citep{2017MNRAS.468..239C}. This will be the focus of a future
  work.

\begin{acknowledgements}
  We  wish  to  acknowledge  the  suggestions  and
   comments of an anonymous referee that strongly improved the original
   version of this work.
  We gratefully acknowledge A. Cumming from providing us with the phase
  diagram and L. Siess for the chemical profiles of his models.  Part
  of this work was supported by AGENCIA through the Programa de
  Modernizaci\'on Tecnol\'ogica BID 1728/OC-AR, and by the PIP
  112-200801-00940 grant from CONICET.  This research has made use of
  NASA's Astrophysics Data System.
\end{acknowledgements}

\bibliographystyle{aa}
\bibliography{paper-one-pulsa.bib}
%\begin{thebibliography}{}
%\bibitem[Auri\`ere(1982)]{aur82} Auri\`ere, M.  1982, \aap,
%    109, 301
%\end{thebibliography}

\end{document}